\documentclass{article}
\usepackage{spconf,amsmath,graphicx}
\usepackage{booktabs}
\usepackage{multirow}
\usepackage{url}
\usepackage{amsfonts}

\usepackage{xcolor}

\title{Video Quality Evaluation Methodology and Result of AV2 Compression Performance }
\name{
    Zhijun Lei$^{\dagger}$, 
    Vibhoothi Vibhoothi$^{\parallel}$,
    Dzung Hoang$^{\mathsection}$, 
    Yixin Du$^{\mathsection}$, 
    Ramzi Khsib$^{\P}$ 
    \thanks{Thanks to AOMedia Codec Working Group (CWG) participants, Testing Sub-Group participants, and AOMedia Members.}
}
\address{
    $^{\dagger}$\textit{Meta}, USA, 
    $^{\mathsection}$\textit{Apple}, USA, 
    $^{\P}$\textit{Amazon}, USA \\
    $^{\parallel}$Sigmedia Group, Dept. of Electronic and Electrical Engineering, \textit{Trinity College Dublin}, Ireland \\
    \small 
    $^{\dagger}$ryanlei@meta.com,
    $^{\parallel}$vibhoothi@tcd.ie,
    $^{\mathsection}$\{dzung\_hoang,yixin\_du\}@apple.com,
    $^{\P}$khsramzi@elemental.com
}

\begin{document}
%
\maketitle
\begin{abstract}
The Alliance for Open Media (AOMedia) has developed the AV2 video coding standard to supersede AV1, aiming for substantial compression efficiency gains across diverse media applications. This paper details the quality and performance evaluation methodology defined in the AV2 Common Test Conditions (CTC), which introduces new evaluation methods and content, including convex-hull-based adaptive streaming (AS) configuration, user-generated content (UGC), and extended chroma formats. We present the coding gains of the AV2 (v13.0) against the AV1 baseline. Experimental results show that AV2 achieves significant Bj{\o}ntegaard-Delta Rate (BD-rate) reductions of 29.81\% and 33.79\% for PSNR-YUV and VMAF, respectively, under random access configuration, validating the efficiency of AV2 for next-generation streaming applications.
\end{abstract}
\section{Introduction}
\label{sec:intro}
To address the increasing demand for efficient video storage and transmission, the Alliance for Open Media (AOMedia) is developing AV2, the next-generation video coding standard succeeding AV1~\cite{av1paper}. 
AV2 aims to deliver a substantial reduction in bitrate compared to AV1 for equivalent visual quality. This efficiency gain is essential for reducing storage and distribution costs, improving video quality in bandwidth-constrained environments, and supporting demanding applications such as UHD-2 (8K) streaming, virtual reality (VR), and augmented reality (AR).

AV2 is being developed with a strong emphasis on practical deployment. The standard sets clear targets for computational complexity, aiming for no more than a five times increase in encoder complexity and a two times increase in decoder complexity relative to AV1 with production-ready implementations. This balance ensures that AV2 can be deployed across a wide range of devices, from high-end servers to mobile and embedded platforms. Furthermore, AV2 continues the tradition of open-source reference implementations of AOMedia, fostering broad industry adoption and innovation.

The development of the AV2 video coding standard requires a rigorous, transparent, and reproducible framework for evaluating new coding tools and overall codec performance. To address this need, the AOM Testing Sub-Group established the AV2 CTC~\cite{aomctc}, a standardized set of guidelines, datasets, and methodologies that serve as the official benchmark for all technical proposals and progress assessments. 
By mandating that all contributors use the same test conditions, the CTC ensures that results are directly comparable, fair, and objective. This uniformity is essential for accurately tracking improvements in compression efficiency, video quality, and computational complexity as new tools are proposed and integrated into the reference implementation. 
By ensuring result reproducibility, the CTC facilitates objective, data-driven decision-making regarding the adoption of coding tools for the AV2 coding standard.


This paper provides a high-level overview of the AV2 CTC design and the latest AV2 performance test results. This paper is organised as follows. Background on AV2 testing condition is presented in Section~\ref{sec:background}, an overview of quality and performance evaluation methodology in the AV2 CTC is presented in Section~\ref{sec:testmethod}, and the latest test results are detailed in Section~\ref{sec:results}. Finally, the conclusion is given in Section~\ref{sec:conclusion}.

\section{Background}
\label{sec:background}
Developing a standardized performance evaluation is the fundamental part of video coding standard development. It allows for the rigorous assessment of coding tools, ensuring that compression gains justify the associated computational complexity. Historically, groups such as MPEG and VCEG, and Joint Video Experts Team  JVET have established CTCs ~\cite{hevcctc, jvetsdrctc, jvethdrctc} primarily focused on broadcast and storage applications. However, the AOMedia operates with a distinct focus on the open web and streaming video, necessitating a unique testing philosophy for the AV2 development cycle.

Unlike traditional CTCs, which often rely on a limited set of four quantization parameters (QPs), the AV2 CTC mandates the use of six QP points for mandatory test classes. This expanded range mitigates interpolation errors when constructing rate-distortion (RD) curves and allows for finer grained analysis across low, medium, and high bitrate ranges using piecewise cubic Hermite interpolation.

Furthermore, AV2 CTC~\cite{aomctc} introduces distinct features which are novel in the CTC testing. Firstly, the CTC includes UGC~\cite{2019youutube-ugc-dataset} that is comprised of content with varying quality and pre-existing compression artefacts. Doing so recognizes the dominance of social media platforms (e.g, YouTube, Facebook, TikTok). Secondly, CTC introduces adaptive-streaming (AS) coding configuration to simulate real-world video-on-demand (VoD) pipelines. The CTC defines a convex-hull-based evaluation method ~\cite{2018_spie_dyn_opt, convexhullmethod} where sequences are encoded across multiple resolutions, and quality metrics are calculated against the original source after upsampling. Lastly, the CTC includes an evaluation for dynamic resolution changes within a sequence (resize mode), allowing the codec to switch resolutions at frame boundaries without IDR frames, a critical feature for bandwidth adaptation.

While PSNR remains a standard metric in testing, the AV2 CTC emphasizes perceptual quality through the mandatory reporting of PSNR-HVS~\cite{egiazarian2006new_psnrhvs}, weighted PSNR (YUV) where we apply different weights per plane (4:2:0: 7/8, 1/16, 1/16; 4:2:2: 4/5,1/10,1/10, 4:4:4:  2/3, 1/6, and 1/6), SSIM, MS-SSIM~\cite{msssimpaper}, CIEDE2000~\cite{yang2012ciede2000}, video multi-method assessment fusion (VMAF)~\cite{vmafpaper}, banding in flat gradients using contrast aware multiscale banding index (CAMBI)~\cite{2021_cambi_paper}. The metrics are implemented inside libvmaf tool (vmaf v3.1.0)~\cite{libvmafurl}.


\section{Quality and Performance Evaluation Methodology in AV2 CTC}
\label{sec:testmethod}
The AV2 CTC defines the specific video sequences, the encoding and decoding configurations, the quality and complexity metrics, and the procedures for conducting coding gain evaluations. 
It ensures a fair comparison can be made between different proposed coding tools and the existing baseline. 
It also helps to evaluate the trade-off between compression gain and complexity, ensuring the codec remains implementable. 

\subsection{Selection of Test Video Sequences}
At the beginning of the AV2 development cycle, a diverse set of source videos and images were selected as the dataset for coding tool evaluation. It covers various resolutions (from 270p up to 4K/UHD), content types (natural videos, synthetic content, and user-generated videos) and formats (SDR with BT.709-BT.1886, HDR with BT.2100 with PQ EOTF, and extended chroma formats like 4:2:2 and 4:4:4), which are grouped into multiple classes. 

By using a wide range of test sequences and scenarios, the CTC helps to ensure that new coding tools perform well with a wide variety of content rather than just a narrow sampling of videos. All videos in each class have the same colour subsampling (except Class ECF), and are encoded with the same number of frames. In addition, all test videos are publicly available\footnote{\url{https://media.xiph.org/video/aomctc/test_set}} to allow reproducibility of results. The test sequences include:
\begin{enumerate}
    \item Class A: natural content, 270p to 2160p, 8/10 bit, 4:2:0;
    \item Class B: synthetic content for gaming, animation, and screen sharing, 1080p, 8/10 bit, 4:2:0;
    \item Class E: user-generated content (UGC) up to 4K, 8-bit, 4:2:0;
    \item Class F: still image up to 8K, 8 bit, 4:2:0;
    \item Class G: HDR in BT.2100 colour space with PQ transfer function, 2160p, 10bit, 4:2:0;
    \item Class ECF: extended chroma format up to 4K, 8/10 bit with total of six sub-classes, 4:2:2 and 4:4:4 with SDR and HDR-BT.2100-PQ, and separate sub-classes for RGB, YCoCg-RE.
\end{enumerate}

\subsection{Encoding Configuration}
Since the main goal of the CTC design is to evaluate normative coding tools, encoding algorithms such as two-pass or look-ahead encoding, content-adaptive QP modulation, and pre-filtering are considered non-normative algorithms and are not enabled in the CTC. In AV2 CTC, the following four major test configurations are defined.
\begin{enumerate}
    \item \textbf{All Intra (AI)} evaluates intra coding methods. This configuration uses the first 15 frames of the video sequences for all classes except Class ECF and the still image Class F. Class ECF uses the first 5 frames.
    \item \textbf{Random Access (RA)} evaluates video on-demand streaming, one-to-many live streaming, and storage use-case, which allows the use of frame reordering. In total, 130 frames (two closed GOPs) are used, except for class ECF (66 frames). 
    \item \textbf{Low Delay (LD)} targets real-time communications and videoconferencing use cases,  where the codec does not use frame reordering. Only one key frame (frame 0) is used. In total, 130 frames are coded except class ECF (33 frames).
    \item \textbf{Adaptive Streaming (AS)} is introduced in Section~\ref{sec:background}. This evaluates the adaptive bitrate streaming (ABR) use case, which involves generating encodings for multiple resolution to construct a convex hull for rate-distortion analysis. For example, with Class A1 ($3840\times2160$p), five downsampled resolutions are used: 2560$\times$1440p, 1920$\times$1080p, 1280$\times$720p, 960$\times$540p, and 640$\times$360p. Each resolution is encoded using the random access configuration.     
    The decoded lower-resolution sequences are subsequently upsampled to the source resolution to compute quality metrics. The methodology for resampling and constructing the Pareto-frontier (convex hull) across resolutions is detailed in Section~\ref{sec:method:convex-hull}.
\end{enumerate}

To ensure high parallelism during development, the CTC mandates specific frame tiling configurations (e.g., 4 tiles for 4K RA) and threading models for different classes. A full verification test is defined and conducted with all classes in single tile mode after each release by the AOM Testing Sub-Group as an official report of the coding gain improvement, while the development cycle uses a mandatory subset of videos. 

\subsection{Construction of Convex-Hull for Adaptive Streaming}
\label{sec:method:convex-hull}

The AS configuration requires precise resolution conversion to compute the Bj{\o}ntegaard-Delta Rate (BD-rate) against the original source~\cite{bdrate, bdrate_extended}. Resampling is performed using a Lanczos filter ($\alpha=5$) with 14-bit integer precision, utilizing centered phase alignment and boundary padding via sample replication. Chroma sample positioning is content-dependent: Type 0 (vertical) for BT.709/non-HDR, Type 2 (co-located with luma $(0,0)$) for HDR, and center-siting (``JPEG'') for still images. These filtering operations are implemented in HDRTools~\cite{hdrtools}.

\begin{figure}
    \centering
    \begin{tabular}{cc}
        \includegraphics[width=0.45\linewidth]{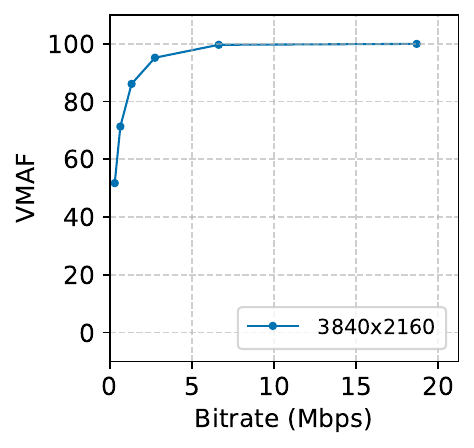} &
        \includegraphics[width=0.45\linewidth]{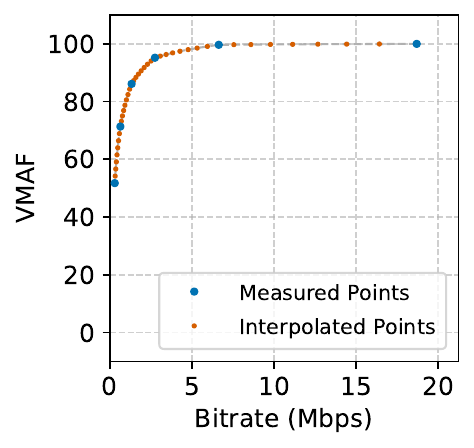} \\
        (a) & (b) \\
        \includegraphics[width=0.45\linewidth]{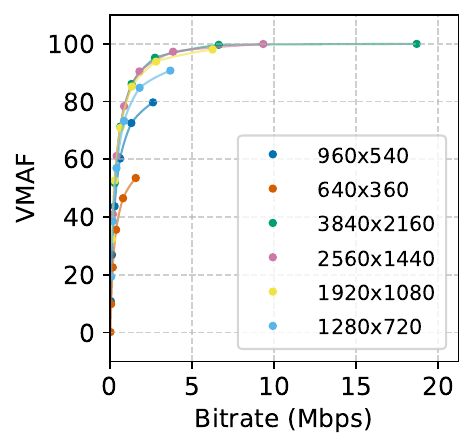} &
        \includegraphics[width=0.45\linewidth]{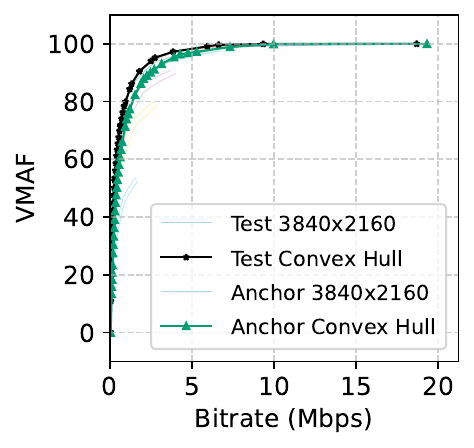} \\
        (c) & (d) 
    \end{tabular}
    
    \caption{
    Step-by-step construction of the convex hull for adaptive streaming evaluation. The process moves from (a) rate-distortion points (VMAF vs. bitrate) for the $3840\times2160$ resolution. (b) log-uniform interpolation between measured points, to (c) multi-resolution aggregation, resulting in (d) the final Pareto-frontier comparison between anchor and test configurations. 
    }
    \vspace{-1em}
    \label{fig:av2-convex-hull-process}
\end{figure}

To ensure sufficient data density for convex hull construction, the six QP points per resolution are interpolated via bilinear interpolation. Seven intermediate points are generated between adjacent QPs, spaced uniformly in the log-bitrate domain. The final convex hull is constructed from the union of measured and interpolated (rate, quality) points across all tested resolutions. Figure~\ref{fig:av2-convex-hull-process} shows this using the reference implementation, convex-hull framework~\cite{convexhull_framework}.

\subsection{Coding Gain Evaluation}
Coding gain is calculated using the Bj{\o}ntegaard rate (BD-rate). This allows the measurement of the bitrate reduction offered by a codec or codec feature, while maintaining the same quality as measured by a given objective quality. The rate change is computed as the average percent difference in rate over a range of qualities. A piecewise cubic Hermite interpolating polynomial (PCHIP~\cite{pchippaper}), which is constrained to be monotonic, is used to fit the RD points. Results are reported for standard PSNR (Y, U, V) and perceptual metrics.

In AV2 CTC, weighted BD-rate is also used, which is defined as:
\begin{equation}
    \begin{aligned}
    \text{BD-rate}_\text{weighted} &= A \times \text{BD-rate}_\text{Y} + B \times \text{BD-rate}_\text{Cb} \\ 
    &+ B \times \text{BD-rate}_\text{Cr},
    \end{aligned}
\end{equation}
where  $A$ = 0.92 ($\frac{23}{25}$), $B$ = 0.04 ($\frac{1}{25}$), which is periodically updated in the CTC evaluation based on luma and chroma RD cost balance.
In the CTC, BD-rate is also computed for different quality ranges such as low quality (QP1--QP4), medium quality (QP2--QP5), and high quality (QP3--QP6). For the metrics which exhibit non-monotonic behaviour for the flat (saturated) region of the RD curve for metrics like VMAF, those values are excluded. 
Please refer to~\cite{aomctc} for more details. 

To evaluate the coding gain achieved with each release of the AVM code base~\cite{libavm_url}, a separate release anchor, \texttt{research-alt-v1-anchor\_r4.0}, is used as the baseline for calculating coding gain. This anchor is based on a code branch of the libaom AV1 reference encoder~\cite{libaom_url}, with additional changes in the scaler quantization design to make sure encoding bitrate can reach a much lower range. 

All simulations use the .obu (Open-Bitstream Unit, analogous to Network Abstraction Layer, NAL Units in HEVC/VVC) format as a raw bitstream output. Encoding bitrate is calculated: 
\begin{equation}
    \text{Bitrate}_{\text{kbps}} = \frac{\text{FileSize} \times 8 \times \frac{\text{fps}_{\text{num}}}{\text{fps}_{\text{denom}}}}{\text{FrameNum} \times 1000},
\end{equation}
where fps\_num and fps\_denom are the numerator and denominator of frame-rate (e.g, 60000/1001 for 59.94 fps). Six decimal points are used to match the precision of quality metrics. 

Table~\ref{tab:qindex_values} specifies the QP\footnote{QP is mapped to the qindex syntax element of AV1 and AV2.} values used for the AV2 CTC encoding configurations, which are explicitly applied for each frame type without considering the characteristics of the input video. 
For the all intra and still image configurations, a lower set of QP values is used, which matches the production scenario that higher encoding quality is needed for still images and intra frames in a video sequence.


\begin{table}
    \centering
    \caption{QP for different encoding configurations}
    \label{tab:qindex_configurations}
    \begin{tabular}{@{}ll@{}}
        \toprule
        Configuration & QP Values \\
        \midrule
        Still Image             & 60, 85, 110, 135, 160, 185 \\
        All Intra (AI)          & 85, 110, 135, 160, 185, 210 \\
        Random Access (RA)      & 110, 135, 160, 185, 210, 235 \\
        Low Delay (LD)          & 110, 135, 160, 185, 210, 235 \\
        Adaptive Streaming (AS) & 110, 135, 160, 185, 210, 235 \\
        \bottomrule
    \end{tabular}
    \vspace{-1em}
    \label{tab:qindex_values}
\end{table}


\section{AV2 Performance Test Results}
\label{sec:results}
Over the AV2 development cycle, periodic tests were conducted across all video classes to gauge compression progress. To ensure maximum coding gain, a single-tile configuration is used for the AV2 v13.0 release evaluation for all classes except Class ECF, where AV2 CTC with parallel tiles is used.

Table \ref{tab:overall_ctc_result} summarises the BD-rate savings against the \texttt{research-alt-v1-anchor\_r4.0} baseline for different objective metrics. 
In the AI configuration, AV2 achieves a bitrate reduction of 22.32\% for PSNR-YUV and 23.58\% for VMAF on the overall 4:2:0 dataset. Notably, coding gains are higher for screen content (Class B2), reaching 42.62\% in PSNR-YUV, indicating efficient handling of synthetic content. 
The random access configuration has the highest efficiency gain with an overall reduction of 29.81\% in PSNR-YUV and 33.79\% in VMAF. Similar to the all intra results, screen content in RA mode yields substantial gains ($\approx$40\%), while HDR sequences yield a 31.99\% reduction. 
On analysing for latency-sensitive applications, the LD configuration maintains competitive performance, achieving a 26.05\% bitrate saving in PSNR-YUV and 27.28\% in VMAF. Screen content remains a standout performer in this mode, exceeding 44\% reduction, which is critical for real-time screen sharing scenarios. 
While for the AS configuration, where a convex hull is constructed across multiple resolutions, AV2 delivers a 31.34\% gain in PSNR-YUV and 35.77\% in VMAF. 
Furthermore, still image (Class F) coding also benefits from inter-coding improvements of AV2, resulting in a 15.12\% bitrate reduction when measured with VMAF.

Evaluation of AV2 with extended chroma format (ECF) tests indicates consistent gains for 4:2:2 and 4:4:4 content, with overall PSNR-YUV savings of 25.40\%, 32.08\%, and 28.75\% for AI, RA, and LD modes, respectively. A consistent trend across all configurations is that chroma component gains generally exceed those of the luma plane, contributing to higher PSNR-YUV scores relative to PSNR-Y. Additionally, the superior compression of screen content suggests that new tools of AV2 are highly effective for computer-generated graphics and sharp edge preservation.

\begin{table*}[t]
\centering
\small
\caption{BD-rate gains for different coding configurations against different objective metrics with CTC and AV2 v13.0. Bold indicates overall performance for different coding configurations.}
\label{tab:overall_ctc_result}
\resizebox{0.87\textwidth}{!}{%
\begin{tabular}{@{}clcccccc@{}}
\toprule
\multirow{1}{*}{Coding Config.} & \multirow{1}{*}{Summary} & \multirow{1}{*}{PSNR-Y} & \multirow{1}{*}{PSNR-YUV} & \multirow{1}{*}{SSIM} & \multirow{1}{*}{MS-SSIM} & \multirow{1}{*}{VMAF} & \multirow{1}{*}{CIEDE2000} \\\midrule
\multirow{10}{*}{\begin{tabular}[c]{@{}c@{}}All\\  Intra\end{tabular}}  & Class A+B1 & 17.82\% & 19.75\% & 18.59\% & 16.87\% & 21.26\% & 25.78\% \\
 & Class B2 (SCC) & 40.87\% & 42.62\% & 45.70\% & 42.16\% & 40.89\% & 50.89\% \\
  & Class G (HDR) & 20.17\% & 24.14\% & 22.10\% & 18.61\% & 22.31\% & 34.57\% \\
 & Class E (UGC) & 16.65\% & 18.49\% & 16.28\% & 15.53\% & 22.70\% & 21.07\% \\ 
 & ECF YCgCo & 18.08\% & 24.20\% & 20.40\% & 18.98\% & 17.07\% & 35.89\% \\
 & ECF SCC & 37.06\% & 38.39\% & 43.03\% & 36.65\% & 37.98\% & 41.24\% \\
 & ECF 4:4:4 & 19.52\% & 22.63\% & 20.41\% & 18.75\% & 21.41\% & 25.55\% \\
 & ECF 4:2:2 & 18.65\% & 21.05\% & 20.18\% & 17.87\% & 20.90\% & 23.46\% \\ \cmidrule(l){2-8}
 & \textbf{ECF Overall} & \textbf{22.92\%} & \textbf{25.40\%} & \textbf{25.12\%} & \textbf{22.23\%} & \textbf{24.74\%} & \textbf{28.12\%} \\
 & \textbf{4:2:0 Overall} & \textbf{20.18\%} & \textbf{22.32\%} & \textbf{21.29\%} & \textbf{19.35\%} & \textbf{23.58\%} & \textbf{28.50\%} \\ \midrule
 \multirow{10}{*}{\begin{tabular}[c]{@{}c@{}}Random\\  Access\end{tabular}} & Class A+B1 & 27.32\% & 28.82\% & 27.20\% & 26.39\% & 32.95\% & 33.03\% \\
  & Class B2 (SCC) & 39.42\% & 40.63\% & 42.05\% & 41.05\% & 43.04\% & 47.05\% \\
 & Class G (HDR) & 29.28\% & 31.99\% & 29.53\% & 27.82\% & 37.08\% & 39.18\% \\
 & Class E (UGC) & 23.73\% & 25.63\% & 23.06\% & 22.49\% & 29.30\% & 28.59\% \\
 & ECF YCgCo & 28.65\% & 32.18\% & 28.85\% & 27.52\% & 28.64\% & 40.18\% \\
 & ECF SCC & 36.71\% & 37.77\% & 39.58\% & 37.21\% & 40.40\% & 40.50\% \\
 & ECF 444 & 28.72\% & 30.83\% & 27.92\% & 26.90\% & 32.38\% & 32.65\% \\
 & ECF 422 & 28.32\% & 30.23\% & 27.94\% & 27.58\% & 34.68\% & 32.86\% \\ \cmidrule(l){2-8}
  & \textbf{ECF Overall} & \textbf{30.27\%} & \textbf{32.08\%} & \textbf{30.40\%} & \textbf{29.34\%} & \textbf{34.92\%} & \textbf{34.39\%} \\
 & \textbf{4:2:0 Overall} & \textbf{28.13\%} & \textbf{29.81\%} & \textbf{28.22\%} & \textbf{27.32\%} & \textbf{33.79\%} & \textbf{34.38\%} \\ \midrule
\multirow{7}{*}{\begin{tabular}[c]{@{}c@{}}Low\\ Delay\end{tabular}}  & Class A+ B1 & 21.25\% & 22.98\% & 21.35\% & 20.12\% & 23.89\% & 27.99\% \\
 & Class B2 (SCC) & 41.04\% & 42.07\% & 42.66\% & 41.85\% & 44.97\% & 46.94\% \\
 & ECF YCgCo & 22.78\% & 26.79\% & 24.40\% & 23.25\% & 26.15\% & 30.45\% \\
 & ECF SCC & 35.93\% & 37.09\% & 38.79\% & 35.65\% & 38.24\% & 40.03\% \\
 & ECF 444 & 24.67\% & 26.77\% & 24.73\% & 23.47\% & 26.34\% & 28.21\% \\
 & ECF 422 & 22.13\% & 23.69\% & 21.25\% & 21.27\% & 24.96\% & 27.05\% \\
 \cmidrule(l){2-8}
 & \textbf{ECF Overall} & \textbf{27.11\%} & \textbf{28.75\%} & \textbf{25.45\%} & \textbf{27.67\%} & \textbf{24.36\%} & \textbf{29.35\%} \\
 & \textbf{4:2:0 Overall} & \textbf{24.43\%} & \textbf{26.05\%} & \textbf{24.78\%} & \textbf{23.61\%} & \textbf{27.28\%} & \textbf{31.03\%} \\ \midrule
\begin{tabular}[c]{@{}c@{}}Adaptive\\ Streaming\end{tabular} & \textbf{4:2:0 Overall} & \textbf{30.25\%} & \textbf{31.34\%} & \textbf{30.59\%} & \textbf{28.11\%} & \textbf{35.77\%} & \textbf{34.74\%} \\ \midrule
Still Image & \textbf{Class F} & \textbf{12.07\%} & \textbf{13.81\%} & \textbf{12.48\%} & \textbf{12.51\%} & \textbf{15.12\%} & \textbf{16.84\%} \\ \bottomrule
\end{tabular}%
}
\vspace{-0.5em}
\end{table*}

\begin{figure}[t!]
    \centering
    \includegraphics[width=0.87\linewidth]{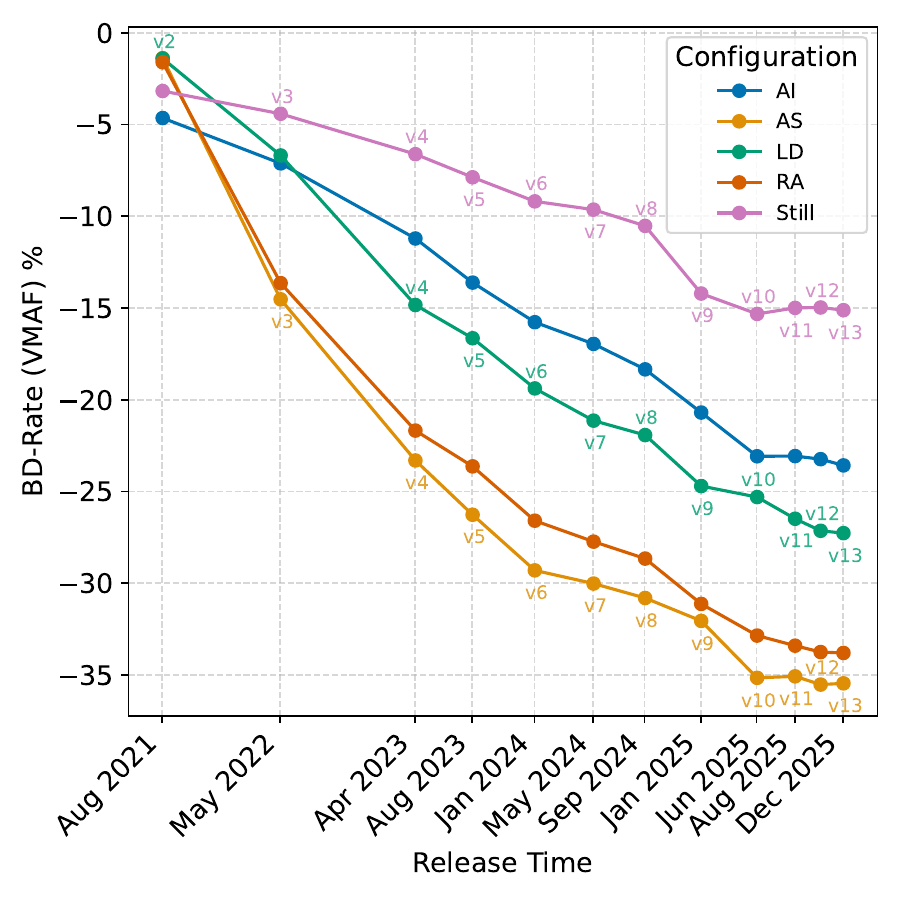}
    \caption{Progress of AV2 Encoder development (2021 to 2025) measured using BD-rate of VMAF (lower is better) for different coding configurations (each line) of AV2 vs AV1 (from v1 to v13 research anchors, circle dots). }
    \vspace{-0.5em}
    \label{fig:av2-coding-progress-vmaf}
\end{figure}

\begin{figure}[t!]
    \centering
    \includegraphics[width=.87\linewidth]{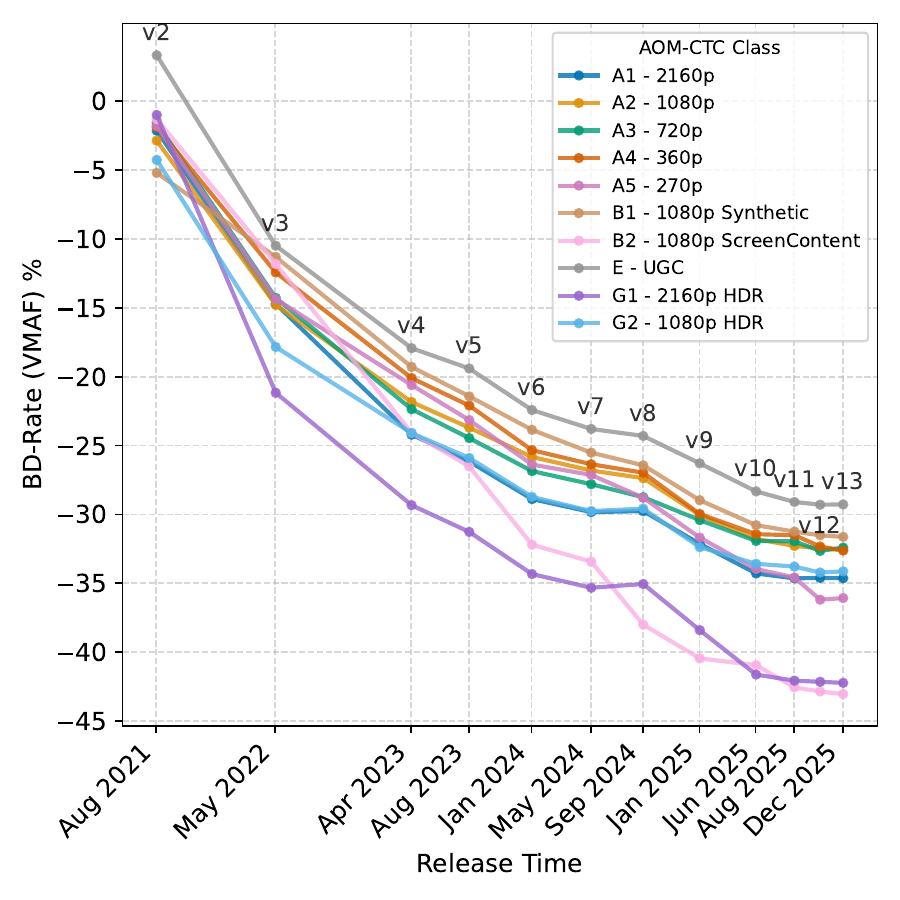}
    \caption{Progress of AV2 Encoder development (2021 to 2025) measured using BD-rate of VMAF (lower is better) for different AV2 CTC classes (each line) of AV2 vs AV1 (from v1 to v13 research anchors, circle dots) in RA Configuration. }
    \vspace{-1em}
    \label{fig:av2-coding-progress-vmaf-by-class}
\end{figure}

\subsection{Coding Gain Progress overall past releases}
Throughout AV2 development, periodic tests monitored compression efficiency after each AVM release. Figures \ref{fig:av2-coding-progress-vmaf} and \ref{fig:av2-coding-progress-vmaf-by-class} track VMAF BD-rate savings across configurations and Test Classes from 2021 to 2025. 
Coding gains improved steadily, with RA and AS configurations showing the largest bitrate reductions. AV2 achieved over 40\% VMAF BD-rate savings for 2160p HDR, and $\approx$30\% for UGC.
Notably, the rate of improvement plateaus post-2025, reflecting the increasing challenge of achieving additional efficiency gains as the standard matures. 
On average, AV2-CTC encoding and decoding take 33$\times$ and 3.86$\times$ longer than AV1 anchor, varying by resolution. While optimizations continue, faster presets already achieve 10$\times$ faster encoding speed with $\le$2\% coding loss.

\section{Conclusion}
\label{sec:conclusion}
This paper presents the AV2 CTC, the methodology established to evaluate the AV2 video coding standard. 
With standardized set of test sequences, encoding configurations, and objective quality metrics, the AV2 CTC ensures fair, transparent, and reproducible comparisons of coding tool proposals and codec releases throughout the development cycle. 
Experimental results show that AV2 achieves a bitrate reduction of 33.79\% and 35.77\% in VMAF for random access and adaptive streaming configurations, respectively, over the AV1 baseline.
Notably, screen content coding achieved $\approx$40\% savings.
Although the current decoding complexity is 2.5$\times$ to 4$\times$, active optimizations is ongoing to deliver a production ready software decoder~\cite{dav2d_project_page_2026}.

\bibliographystyle{IEEEbib}
\bibliography{refs}

\end{document}